# BEAM BASED UNDULATOR TRAJECTORY ALIGNMENT AT EUROPEAN XFEL*

M. Scholz†, F. Brinker, W. Decking, L. Froehlich, Deutsches Elektronen-Synchrotron DESY, Hamburg, Germany
W. Freund, European XFEL GmbH, Schenefeld, Germany

*Abstract*

Precise alignment of the undulator trajectory onto a straight line is a per-requisite for high fidelity SASE operation, which in turn also enables non-standard operation modes like self-seeding or two-color operation. At European XFEL electron and photon beam-based method are combined with a step-by-step performance-based optimization to ensure that the trajectory is on a straight line over most of the length of the about 250 m long undulator system.

## INTRODUCTION

The undulator lines of European XFEL consist of up to 37 unit cells providing space for a 5 m long insertion devices each, this could be an undulator or a delay/monochromator chicane, and a 1.1 m intersection hosting a synchrotron radiation absorber and a pump port, a cavity beam-position-monitor (BPM), a quadrupole that can be moved remotely in horizontal/vertical direction, and a permanent magnet phase shifter (see Fig. 1). Each yaw of the planar permanent magnet undulators can be moved separately in vertical direction, thus providing control over the vertical undulator offset.

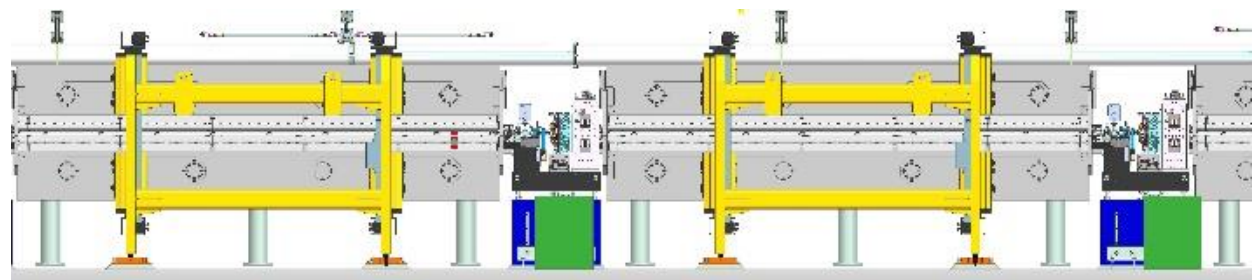

Figure 1: CAD drawing of two undulator cells with (from left to right) a 5 m long undulator and the 1.1 m intersection.

The challenge for SASE FELs with long gain length is the proper alignment of the undulator system so that the radiation field of the undulator overlaps with the electron trajectory. Typcial accuracy requirements are in the micron resp. sub-µrad level.

Quadrupole and BPM misalignment, residual field errors from undulators and phase shifters, and the ambient field make it impossible to achieve this accuracy without beam based methods. Quadrupoles and BPMs are initially aligned with geodetic methods and we assume a conservative precision of about 500 micron P-P. With a typical field gradient of 20 T/m we come to 1 Tmm P-P. Undulators and phase shifter magnetic fields have been precisely measured in the laboratory. Undulators can exhibit an entrance and exit kick that as large as 0.15 Tmm RMS. This kick is gap dependent. Based on accurate streteched wire measurements the first and second field integrals of the undulators are corrected with the help of air-coils at the beginning and exit of the unudulator to values to well below 0.01 Tmm resp. 0.1 $Tmm^2$ [1]. These values are only achieved by ‘subtracting’ large numbers (i.e. the undualtor kick and the air-coil kick from the undualtor kick) and thus have the potential for large trajectory deviations should for instance the air-coil not track the undulator gap correctly. The kick from the phase shifters is also gap dependent and smaller than 0.004 Tmm RMS for gaps larger than 16 mm, and up to 0.016 Tmm RMS for the smallest 10 mm gap.

Apart from being measured in the magnetic laboratory, residual kicks from phase shifters and undulators can also be measured with beam as a difference trajectory measurement for various gaps. These measurements largely confirm the numbers given in Tab. 1, with some excpetions for undulators where the measured kick can be an order of magnitude larger. The air coil correction is than further improved based on the beam based measurements [2].

Table 1: Field Errors in undulator section

| Object | Kick [Tmm] |
|---|---|
| Undulator entrance | 0.15 RMS |
| + air coil | 0.01 RMS |
| Undulator exit | 0.15 RMS |
| + air coil | 0.01 RMS |
| BPM, absorber (earth field) | 0.01 (in y) |
| Quadrupole | 1 P-P |
| Phase Shifter (> 16 mm gap) | 0.004 RMS |

## VERTICAL UNDULATOR OFFSET ALIGNMENT WIT ELECTRON BEAM BASED ALIGNMENT OF QUADRUPOLES AND BEAM POSITION MONITORS

Initial alignment of the quadrupole and BPM offsets is performed by the established dispersion free steering method [3]. It’s application at European XFEL has been described in [4]. The process has been improved over the past years and nowadays terminates after an almost dispersion free trajectory is reached (≈ 1 µm/GeV) and the predicted variation of the quadrupole position for the next step is below 5 µm.

* Work supported by operation funds of European XFEL
†matthias.scholz@desy.de

Compared to the initial geodetic alignment the quadrupoles have been moved up to 500 µm (see Fig. 2). Once the Electron BBA is finished, the trajectory is straight enough for the following steps that fine tune the undulator trajectory alignment.

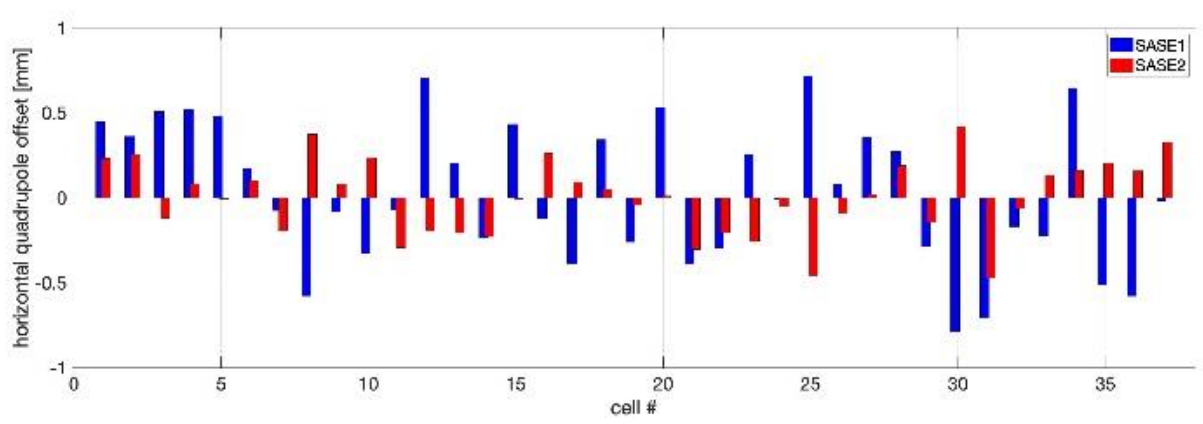

Figure 2: Horizontal quadrupole offsets after electron beam-based alignment.

## PHOTON-BASED ALIGNMENT OF SINGLE UNDULATOR POINTING

Given the residual fields as summarized in Tab.1, the trajectory in between two quadrupoles/BPMs can still have an angle with respect to the overall straight line of up to 0.5 µrad. Also, the accuracy of the electron beam-based alignment gets worse towards the end of the undulator, where less BPMs are involved, and at the beginning of the undulator, where the alignment is dominated by the incoming trajectory. For these cases the photon-based alignment also fixes the trajectory angle in each undulator.

This measurement can be performed with the help of a K-monochromator ("K-mono"). As discussed in [5], this monochromator only lets a very narrow spectral range around a selectable wavelength pass, and we usually use it to inspect the radiation emitted by a single undulator cell on a screen downstream of the K-mono. If the monochromator wavelength $\lambda_{mono}$ is slightly detuned from the resonance wavelength, we see how the off-axis radiation forms a ring (Fig. 3) whose radius is determined by the opening angle $\theta$ of the undulator radiation for this particular wavelength.

By comparing a tuned and a de-tuned undulator cell to each other one can easily determine the centre of the on- and off-axis radiation and align the pointing to coincide using the air-coils around one of the undulator. Repeating this procedure for all cells in principle aligns the pointing of all undulators.

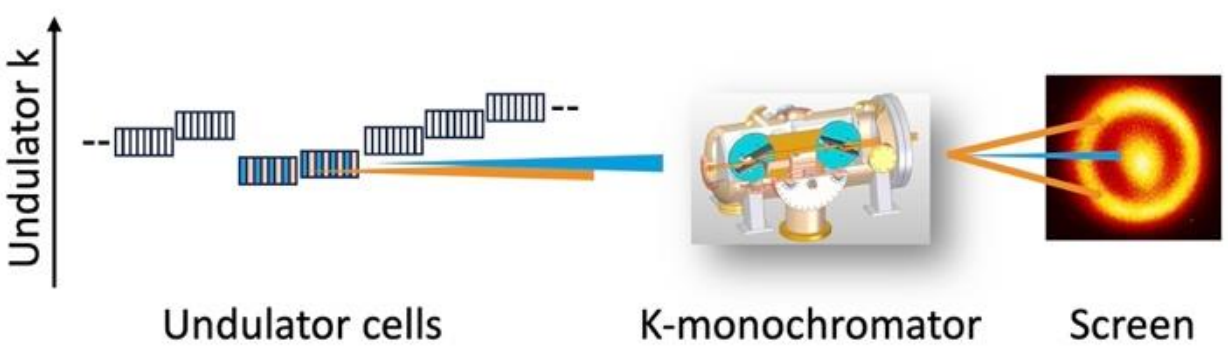

Figure 3: Schematic depiction of the photon-based alignment using the K-monochromator.

## VERTICAL UNDULATOR OFFSET ALIGNMENT WITH K-MONOCHROMATOR

Apart from making the trajectory as straight as possible, it is also important to centre the beam well enough inside the undulators. The good field region in the horizontal plane is with 1.8 Tmm relatively broad, so that horizontal offsets are usually not a concern. In the vertical plane, however, the amplitude of the undulator field varies as

$$B_y = B_0 \cosh {}^{2\pi}/_{\lambda_p}\, y \tag{1}$$

where $\lambda_p$ is the undulator period, $B_0$ the amplitude of the magnetic field on axis, and $y$ the vertical distance from the axis [6]. In other words, the magnetic field seen by the beam increases with increasing vertical distance from the axis, and so does the effective undulator parameter K. This leads to a shift towards longer wavelengths.

We can measure this effect with the help of the K-mono. Using the well-known undulator formula

$$\lambda_{mono} = \lambda_p \frac{1}{\gamma^2 2n}\left(1 + \frac{1}{2}K_{eff}^2 + \gamma^2\theta^2\right) \tag{2}$$

with the relativistic Lorentz factor $\gamma$ and the number n of the harmonic under observation, it is straightforward to calculate the effective undulator parameter $K_{eff}$ seen by the beam from the off-axis radiation observed with the K-mono.

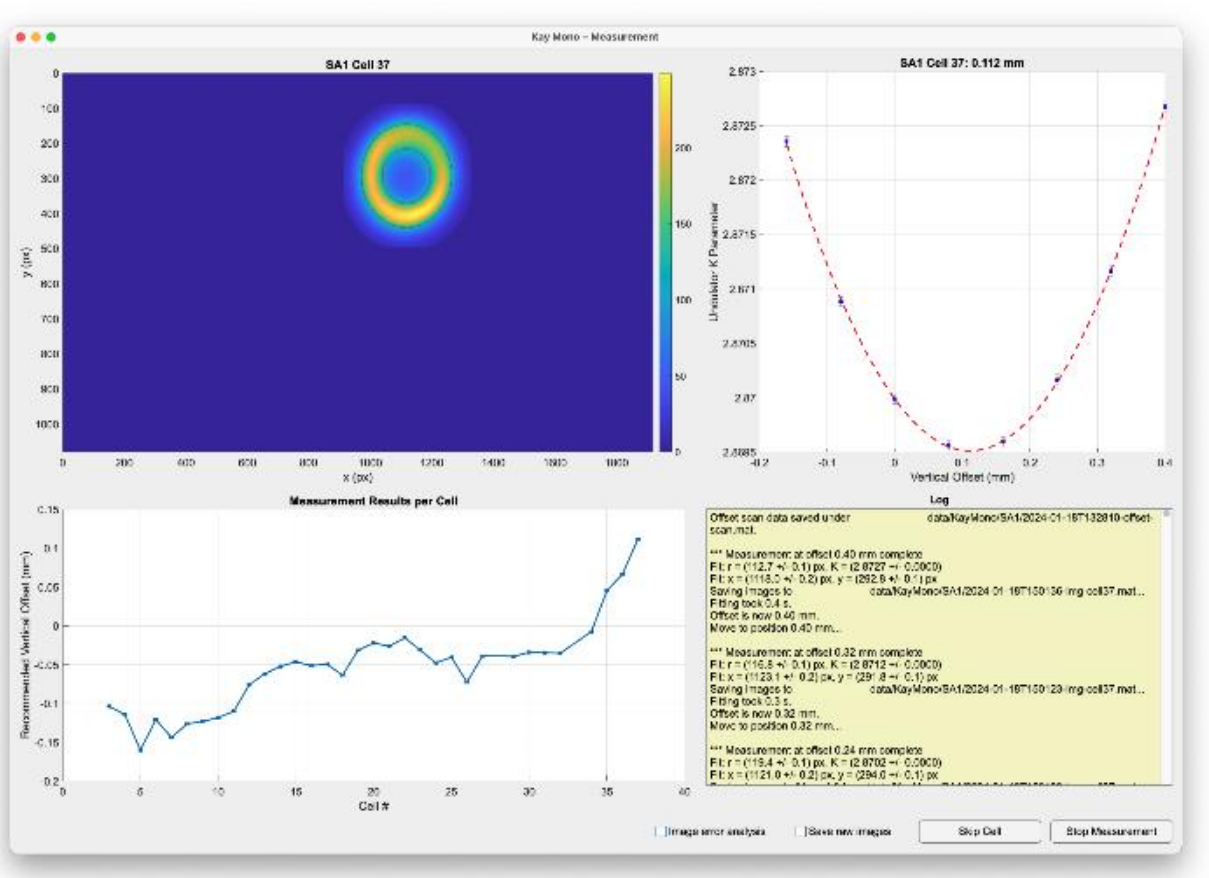

Figure 4: Screenshot of a K-Mono measurement with the screen image after the monochromator top left, a fit of Eq. (1) to the radii determined for various vertical offsets top right, and the results for all undulator cells of the SASE1 undulator bottom left.

We occasionally use the K-mono as a tool for aligning the undulators vertically around the trajectory that has been established by the BBA. For this, we fix the electron trajectory with an orbit feedback and close only one undulator cell while all others are opened. Both undulator jaws are now moved upwards or downwards in lockstep, keeping the gap constant.

This is repeated for a small number of vertical offsets. At each point, a fitting algorithm determines the radius of the observed ring, and the effective K is calculated (Fig. 4 top right). As a result, we obtain a curve as predicted by Eq. (1) and can determine the ideal offset, i.e. the vertical offset at which the undulator jaws are centered around the electron beam. If these offsets are sufficiently large, the undulators are permanently set to the new vertical position.

## SLIDING WINDOW OPTIMIZATION

The sliding window optimization (SWO) method is a process used to verify and fine-tune the alignment of the hard X-ray undulator beamlines on a section-by-section basis. This is achieved by considering the intensity and pointing of the photon beam [7].

This approach is typically performed as the final optimization step following any electron energy changes. Although this step is typically applied to both hard and soft X-ray beamlines, this description will focus on the beamlines for high photon energies, as these are more sensitive to misalignment and therefore benefit more from the method.

The typical parameters for the SWO approach at European XFEL are an electron beam energy of 14 GeV and a photon energy of 9 keV, which corresponds to an undulator gap size of approximately 13.5 mm. A linear tapering of the undulator k-values is applied in order to compensate for the energy losses of the electrons due to undulator radiation and wake fields. Moreover, the phase shifters between the undulator cells are configured in accordance with the standard operational parameters at the specified electron and photon energies, thereby ensuring a consistent influence on the trajectory.

The number of closed cells at the beginning of the undulator is initially selected to ensure a good signal-to-noise ratio in the photon detector and to verify that the undulator configuration remains securely within the linear gain mode. At EuXFEL, approximately 13 undulator cells are closed, and a photon pulse energy of around 40-50 uJ is used. The detector signal is then well above the noise threshold of a few uJ [8, 9], suggesting that working with fewer cells at the same time may be feasible. However, experience has shown that aligning too few cells at a time can lead to difficulties defining a straight line, which is an important reason for the effort.

As previously outlined, the procedure begins with the initial undulator cells in a closed position. The remaining cells are then adjusted to a slightly reduced photon energy (corresponding to a larger k-value). Furthermore, a robust negative linear taper is implemented to these cells in order to mitigate lasing. This approach is employed in order to prevent significant alterations in the undulator gap size during the SWO procedure, as these changes would influence the electron's trajectory.

The preliminary phase of the process requires the optimization of the trajectory within the cells that have been closed. This is accomplished through the utilization of a generic optimizer tool, which modifies the kicks from air coil correctors at the beginning and end of each cell until their optimal configuration is identified. The resulting pulse energy will serve as the reference point for subsequent optimization stages. If necessary, the pointing of the undulator can still be corrected at this point.

In the subsequent phase, the undulator trajectory feedback is prepared. For this purpose, the beam position monitors located in the central region of the closed undulator cells must be selected. It is important that the first and last cells are not included in the feedback area, as these will be changed in the following step and would provoke an unwanted reaction from the system. Correctors for both planes upstream of the undulator are used as feedback actuators.

Now the actual SWO can begin: The first closed cell must be moved slightly out of resonance, thereby preventing its contribution to the intensity, with minimal impact on the trajectory. Subsequently, the first cell that was not contributing to the intensity at the end of the current active undulator cells is then closed (adjusted to the k-value, including the linear taper of the active cells, see Fig. 5). Should the necessity exist, the trajectory feedback will maintain the electrons in the central area at the same position.

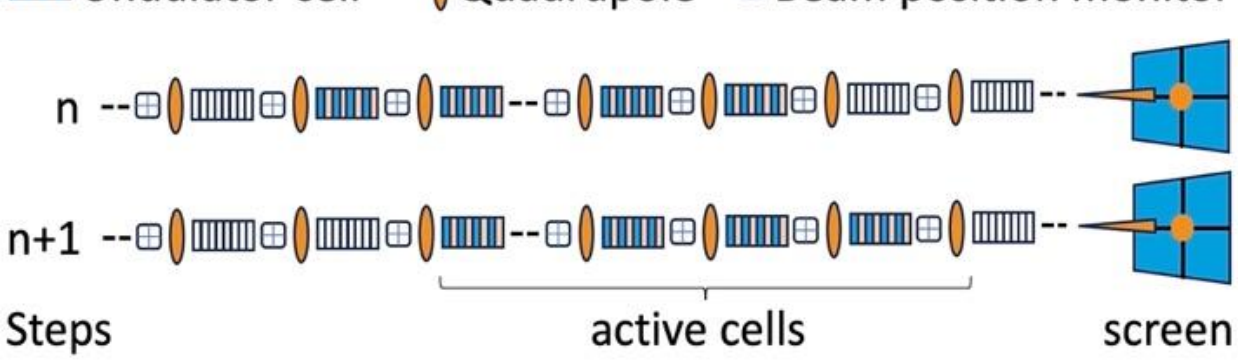

Figure 5: Schematic depiction of the sliding window optimization steps n and n+1. The undulator cells actively involved in the SASE process are moved back one cell after each optimization step.

It is anticipated that, provided the newly closed cell is aligned with the cell that was just opened, the same photon beam position on the imager and the same pulse intensity in the detector will be observed after this step. In the event that both conditions are not met simultaneously, it is necessary to refine the alignment of the final cell. To achieve this, the kicks from the air coil correctors between the last and second-to-last cells are optimized using the optimizer tool.

This procedure is repeated for all remaining undulator cells. At the conclusion of the process, only the last 13 undulator cells remain closed, the pointing of the photon beam is consistent with that of the initial cells, and the radiation intensity is approximately equivalent. Upon achieving these outcomes, it can be concluded that the fine alignment of all cells has been successfully completed. Concurrently, the SWO ensures that all cells contribute equally.

## CONCLUSION

Precise alignment of the undulator trajectory for long undulators and SASE gain length can be achieved by a combination of several methods that partly overcome the intrinsic limitations of each of the individual methods. The complete procedure is time consuming and vulnerable to condition changes throughout the process, and thus will profit from further automation.